\documentclass[prb,twocolumn,aps,showpacs]{revtex4}
\usepackage{graphicx}
\def\be{\begin{equation}}
\def\ee{\end{equation}}
\def\ds{\displaystyle}
\newcommand{\bs}[1]{\rule[3.5mm]{0mm}{0mm}{\bfseries\textsf{#1}}}

\begin{document}

\title{Critical conductance of two-dimensional chiral systems with random magnetic flux}

\author{P. Marko\v{s}$^1$ and L. Schweitzer$^{2}$}%
\affiliation{$^1$Institute of Physics, Slovak Academy of Sciences, 845\,11 Bratislava,
Slovakia\\$^2$Physikalisch-Technische Bundesanstalt, Bundesallee 100, 38116
Braunschweig, Germany}

\begin{abstract}
The zero temperature transport properties of two-dimensional lattice systems
with static random magnetic flux per plaquette and zero mean are investigated
numerically. We study the localization properties and the two-terminal
conductance and its dependence on energy, sample size, and magnetic flux
strength. The influence of boundary conditions and of the oddness of the
number of sites in the transverse direction is also studied. 
For very long strips of finite width, we find a diverging localization length
in the middle of the energy band at $E=0$ and determine its critical exponent
$\nu=0.35\pm 0.03$. A previously proposed crossover from a power-law to a
logarithmic energy dependence can be excluded from our data, at least for
energies $|E| > 10^{-10}$. For square systems, the sample averaged scale
independent critical conductance $\langle g_c\rangle$ turns out to be a
function of the amplitude of the flux fluctuations whereas the variance of the
respective conductance distributions appears to be universal. We find a
critical conductance $\langle g_c\rangle\simeq 1.49\,e^2/h$ for the strongest
possible disorder. 
\end{abstract}


\maketitle

\section{Introduction} 
The transport properties of charged quantum particles in two-dimensional
systems with various types of disorder are of considerable interest in a
variety of experimental and theoretical situations. In particular, the
presence of a static random magnetic flux with zero mean has been of much
concern recently in connection with bond disordered Anderson models with
either real or complex hopping terms,  
\cite{LF81,SN93,Fur99,Cer00,Cer01,ERS01,EK03,GT04,ER04,GC06}
with the composite-fermion picture of the fractional quantum Hall effect at
half-filling,\cite{KZ92,HLR93} 
the critical behavior at the quantum phase transition of spin-split Landau
levels,\cite{LC94} and with the gauge field theory of high-$T_c$
superconductivity.\cite{NL90,AI92}
In addition, far-reaching relations between the low energy chiral limit of
a quantum chromodynamic (QCD) partition function and a large N limit of the
random matrix theory\cite{SV93,AV00} as well as between the electrical
conductance in disordered media and spontaneous chiral symmetry breaking in
QCD have recently become apparent.\cite{GO04,GO07}   
 
Concerning the random flux model, there exists an extensive list of valuable
contributions to this intricate problem (see, e.g., Ref.~\onlinecite{MBF99}
and references therein), but a definite picture started to emerge only
recently, at least for quasi-1d (Q1D) samples.\cite{MBF99}
Results for true two-dimensional systems are scarce and precise numerical
estimates are still missing. 
A consensus has been reached on the notion that all electronic states are
localized for such systems where in addition to the random magnetic flux also
random diagonal disorder is present.\cite{BSK98,PS99,Fur99}   
However, in the absence of diagonal disorder, it has also
been shown that the random flux model with Gaussian distributed and
$\delta$-correlated magnetic fields can be mapped onto a nonlinear
$\sigma$ model of unitary symmetry so that all electronic states should be
localized.\cite{AMW94}
The recognition of a special chiral symmetry that can emerge in systems with
an underlying bi-partite lattice, so that the eigenvalues appear in pairs 
$\pm\varepsilon_i$,\cite{ITA94} has considerably augmented our view of the
possible situations a random flux model can assume.\cite{GW91,Gad93,MW96,BMSA98,AS99}
 
Our aim is to investigate a lattice model with static random magnetic fluxes
and to numerically calculate the two-terminal conductance and the  
localization properties for energies close to the band center. We want to
study the role of the chiral symmetry and to clarify the possible dependence
on boundary conditions (BC). In addition, we address the influence of an odd
or even number of lattice sites. For Q1D systems, we will check the 
assertion that the Lyapunov exponents do not come in pairs for samples with an
even width.\cite{Cer01} We will also look for the crossover proposed for 
the energy dependence of the localization length.\cite{FC00} 
Finally, we calculate the size dependence of the conductance of square
systems and show that at the band center the conductance converges to the
critical value $\langle g_c\rangle \simeq 1.49 e^2/h$. 

The paper is organized as follows.
In Section~\ref{q1d} we study the spectrum of Lyapunov exponents (LE) in the 
quasi-1d limit and for square samples and discuss how the physical symmetry of
the system depends on the boundary conditions, the parity of the width of the
system, and energy of the electron.  In Section~\ref{ce} we find that
$E=0$ is a critical point: the smallest LE does not depend
on the width $L$ of the lattice. For $L$ odd and Dirichlet BC, we also
calculate the critical exponent for the divergence of the localization length
of the two-dimensional system. We find a value $\nu=0.35\pm 0.03$
which is close to the critical exponent for the Anderson bond disordered
model\cite{Cer00,ERS01} and also in agreement with the one obtained with a
different method for the random-flux model which has been reported
recently.\cite{Cer01} 
For Q1D systems of finite width $L$, the localization length diverges
as $\xi\propto L|\ln (|E|L^{1/\nu})|$.
In Section~\ref{cond} we present our data for the critical two-terminal
conductance. Although the scale independent mean value $\langle g_c\rangle$
depends on the strength of the magnetic field fluctuations $f$, the variance
of the corresponding distributions $p_c(g,f)$ turns out to be universal. 
We also confirm the unusual length dependence of the mean conductance for
systems with $L$ odd and DBC.\cite{MBF99} 
Concluding remarks are given in Section~\ref{concl}.

\section{Model and Method}
The two-dimensional (2d) motion of non-interacting particles subject to a
perpendicular random magnetic field is described by a Hamiltonian  
\begin{eqnarray}\label{ham}
{\cal H}&=&
- \,\sum_{m} \Big(t_x (c^\dag_{m+a_x}c^{}_{m}+c^\dag_{m-a_x}c^{}_{m})\\*
&+& t_z(\textrm{e}^{i\alpha_{m,m+a_z}}c^\dag_{m+a_z}c^{}_{m}
+ \textrm{e}^{-i\alpha_{m,m-a_z}}c^\dag_{m-a_z}c^{}_{m})\Big)\nonumber
\end{eqnarray}
with nearest neighbor hopping, defined on the sites $m$ of a 2d square lattice, 
where the width $L$ ($x$-direction) and the length $L_z$ ($z$-direction) 
of the sample is measured in units of the lattice constant $a$. The value of
the hopping term in the $x$-direction is $t_x=1$ if not stated otherwise, 
and the energy is given in units of $t_z=1$.
The operators $c^\dag_{m}$ and $c^{}_{m}$ create or annihilate a Fermi
particle at site $m$, respectively. The complex hopping terms are chosen 
such that the magnetic flux (in units of the flux quantum $h/e$) through 
an individual plaquette is given by the sum of the random Peierls phases 
along the two bonds in the $z$-direction 
$2\pi\phi_m=\alpha_{m,m+a_z}-\alpha_{m+a_x,m+a_z}$. The random
fluxes are distributed uniformly according to $-f/2 \le \phi_m \le f/2$,  
where $0<f\le 1$, with probability density $p(\phi_m)=1/f$ 
so that its second moment is $f^2/12$, and the average magnetic flux 
through the system is zero. The randomness is maximal for $f=1$.  

Without attached leads, the model (\ref{ham}) exhibits chiral unitary symmetry
for Dirichlet boundary conditions in both directions. 
The chirality is destroyed when periodic boundary condition are imposed along
any direction provided the number of sites in this direction is odd.
Table~\ref{symtab} summarizes the various situations.
The chiral symmetry is always broken by an additional on-site disorder.

In the following, we study numerically the quantum transport of electrons
with energy $E$ through the 2d system defined by the Hamiltonian (\ref{ham}). 
For a given length of the system $L_z$, we calculate the dimensionless 
two-terminal conductance via the relation\cite{ES81} 
\be\label{ES}
g=\textrm{Tr}\{T^\dag T\}=\sum_i^{N} \ds{\frac{1}{\cosh^2(x_i/2)}}.
\ee
In Eq.~(\ref{ES}), $T$ is the transmission matrix and the $x_i$ 
parameterize its eigenvalues. The electrons propagate in the 
$z$-direction and $N$ is the number of open channels. 
Dirichlet (DBC) or periodic (PBC) boundary conditions are imposed in the
transversal direction. In the limit of $L_z/L\to\infty$, the parameters $x_i$
converge to the quantities $z_i\times(L_z/L)$,\cite{PZIS90,Mar95a} where $z_i$
is the $i$th Lyapunov exponent (LE) which characterizes the exponential
decrease of the wave function of quasi-1d systems. Oseledec proved\cite{Ose68}
the convergence for the eigenvalues of the transfer matrix, 
$z_i=\lim_{L_z/L\to\infty} z_i(L_z/L)$. For sufficiently large $L_z/L$, the
$z_i(L_z/L)$ are self-averaging quantities. 
The smallest positive LE $z_1$ is related to the localization length and
represents the key parameter of finite-size scaling.\cite{PS81,MK81}
 
\begin{table}
\begin{tabular}{l|c|c|c|c|c|c|c|}
\multicolumn{1}{c}{} & \multicolumn{3}{c|}{$L=$ odd}& &\multicolumn{3}{c}{$L=$ even} \\ \cline{2-8}
\rule[3.5mm]{0mm}{0mm} & &$DBC_x$&$PBC_x$& & &   $DBC_x$&$PBC_x$   \\ \cline{2-4} \cline{6-8}
$L_z=$ odd&$DBC_z$&\bs{CU+}& \bs{U}&  & $DBC_z$&\bs{CU} & \bs{CU}   \\ \cline{2-4} \cline{6-8}
         &$PBC_z$&  \bs{U}  &  \bs{U}&  & $PBC_z$& \bs{U}  &  \bs{U} \\ \cline{1-8}
         & \multicolumn{3}{r|}{} & &\multicolumn{3}{r|}{}\\\cline{1-8}
\rule[3.5mm]{0mm}{0mm} & &$DBC_x$&$PBC_x$&  & &  $DBC_x$&$PBC_x$   \\ \cline{2-4} \cline{6-8}
$L_z=$ even&$DBC_z$& \bs{CU} &  \bs{U} & & $DBC_z$& \bs{CU} & \bs{CU} \\ \cline{2-4} \cline{6-8}
         &$PBC_z$& \bs{CU}  &  \bs{U} & & $PBC_z$& \bs{CU} & \bs{CU} \\ \cline{2-8}
\end{tabular}
\caption{The symmetries of the model Hamiltonian (\ref{ham}) depend on the
  boundary conditions and on the oddness of the number of sites. In the
  absence of leads, the possible symmetry classes are unitary (\textsf{U}),
  chiral unitary (\textsf{CU}), and 
  chiral unitary with an extra eigenvalue that appears at $E=0$
  (\textsf{CU+}).\cite{MW96,BMSA98,AS99,MBF99}}
\label{symtab}
\end{table} 

Since the calculation of the transmission probability requires two semi-infinite 
(ideal in our case) leads attached to the left and right of the sample, 
the boundary condition in the propagation ($z$) direction are neither PBC nor DBC.
We expect, however, that the boundary conditions in the transversal ($x$) direction
affect the transport properties of the system considerably. 

Our data, both for the conductance and for the parameters $x_i$, support 
the conjecture that (\textit{i}) 
the system possesses chiral unitary symmetry only at the band center $E=0$ for
DBC, and for PBC with $L$ even.
The chirality of the $E=0$ state is confirmed by our data for the
parameters $x_i$. In particular, we find that the probability $p(x_1)$
does not decrease to zero when $x_1\to 0$. We will later discuss that this
behavior is typical for the chiral symmetry class. 
(\textit{ii})
There exists a critical point at the band center for $L$ odd and DBC.  
Since this critical point is due to the chiral symmetry of the model, we
expect the criticality also for $L$ even. This expectation is supported by our
numerical data for the smallest LE $z_1$. For $L$ odd and PBC, the critical
state at $E=0$ should disappear due to the unitary symmetry. 

\begin{figure}
\includegraphics[clip,width=0.45\textwidth]{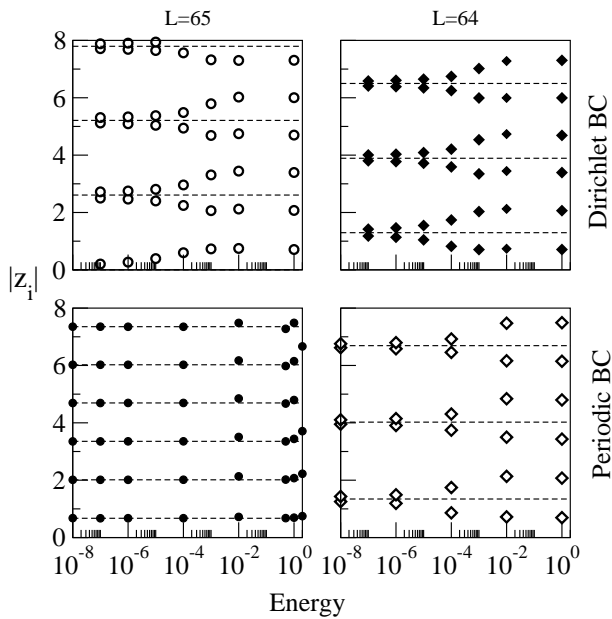}
\caption{The energy dependence of the spectrum of Lyapunov 
exponents $|z_i|$ of the transfer matrix. Dirichlet and periodic 
BC are imposed in the transversal direction, and $f=1$.
Left: $L=65$, right: $L=64$. Dashed lines indicates the values of the Lyapunov
exponents for $E=0$. Note that $z_1=0$ for $L$ odd and Dirichlet BC.
}
\label{figr1}
\end{figure}

\section{Lyapunov exponents}
\label{q1d}
Figure \ref{figr1} shows the spectrum of Lyapunov exponents $|z_i|$
for quasi-1d systems with Dirichlet and periodic BC in the transverse
direction and with either odd ($L=65$) or even ($L=64$) system width.
For $L$ even, the spectrum is degenerate at the band center for both Dirichlet
and periodic BC
\be\label{Leven}
|z_{2i-1}|=|z_{2i}|=c\times \left[i-1/2\right]~~~~~(L~\textrm{even}).
\ee
For $L$ odd, we obtain at the band center that
\be\label{Lodd}
|z_i|=
\ds{\left\{\begin{array}{ll}
c\times \textrm{Int}~\left[i/2\right]   &  (L~\textrm{odd, DBC})\\
~~ & ~~\\
c/2\times\left[i-1/2\right]  &  (L~\textrm{odd, PBC}).
\end{array}
\right.
}
\ee
From Fig.~\ref{figr1} and later from Fig.~\ref{figr8} we see that $c\approx
2.68$.  As is shown in
Fig.~\ref{figr1}, the degeneracy is removed for non-zero energy.
In the transfer matrix method, we calculate only \textit{positive} Lyapunov
exponents. Since the LE appear in pairs, we have also doubly degenerate LE 
($-z_i,-z_i$) in the negative part of the spectra. 

While the form of the spectrum of LE for odd $L$ and PBC is typical for unitary
symmetry,\cite{Pic91} the degeneracy of the spectra, observed in all three other
cases, indicates chiral symmetry.\cite{MW96,BMSA98,MBF99}

For DBC, the chirality is confirmed also by the analysis of the distribution
of parameters $x_i$, calculated for finite length $L_z$.
Since we are able to calculate only the absolute value of the LE, 
we cannot distinguish from the present data whether the system possesses
chiral unitary (\textsf{CU}) or unitary (\textsf{U}) symmetry.
Fortunately, we can estimate the physical symmetry from the analysis of the
distribution of the parameter $x_i$, calculated for systems of finite length
$L_z$.

\begin{figure}
\includegraphics[clip,width=0.38\textwidth]{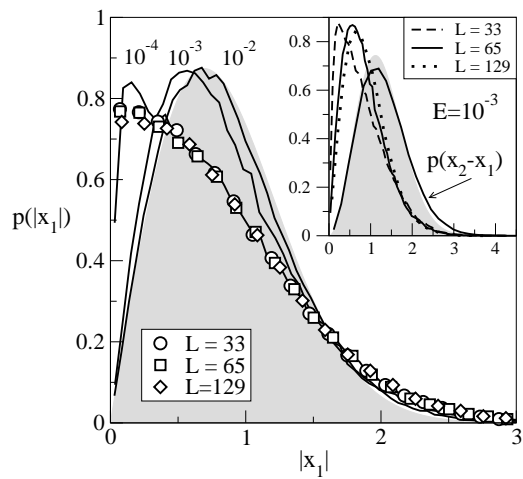}
\caption{The probability distribution $p(|x_1|)$ for $E=0$ and $L=33$, 
65, and 129 (data points). Solid lines are $p(x_1)$ for $L=65$ and
$E=10^{-4}$, $10^{-3}$, and $10^{-2}$ (from the left). The last distribution
is compared with the Wigner surmise 
$\langle x_1\rangle W_1(x_1)=\frac{\pi}{2}s\exp -\frac{\pi}{4}s^2$,
where $s=x_1/\langle x_1\rangle$. The inset shows $p(x_1)$
for $E=10^{-3}$ and $L=33$, 65, and 129. Shown is also the distribution
$p(x_2-x_1)$ for $E=10^{-3}$ and $L=129$ which is almost identical with
the Wigner surmise $\langle x_1\rangle 
W_2(x_1)=\frac{32}{\pi^2}s^2\exp -\frac{4}{\pi}s^2$
for unitary ensemble.
}
\label{figr2}
\end{figure}

As discussed in Refs.~\onlinecite{MBF99} and \onlinecite{BMSA98}, 
for weak disorder the probability distribution $p(\{x\})$ is determined by the  
Dorokhov-Mello-Pereyra-Kumar equation\cite{Dor82a,MPK88} 
\be
\ell\ds{\frac{\partial p}{\partial L_z}}=\ds{\frac{1}{2N}}\sum_{j=1}^N
\ds{\frac{\partial}{\partial x_j}}\left[J\ds{\frac{\partial}{\partial x_j}}(J^{-1}p)\right].
\ee
Here, $\ell$ is the mean free path and $J$ is the Jacobian
\be\label{jacoby}
J=\left\{
\begin{array}{ll}
\prod_{k>j}|\sinh(x_j-x_k)|^2                              &  (\textsf{CU})\\
\prod_{k>j}|\sinh^2 x_j-\sinh^2 x_k|^2 \prod_k |\sin(2x_j)| & (\textsf{U}).
\end{array}
\right.
\ee
The main consequence of the 
absence of the repulsion term $\sin (2x_i)$ in the Jacobian
(\ref{jacoby}) is that the spectrum of $x_i$ spans over the entire
real axis: the $x_i$ can be both positive and negative when the system
possesses chiral unitary symmetry.
In the ordinary unitary systems, all values of $x_i$ are positive, being
reflected from the origin by an additional term in the Jacobian.
Clearly, in the case of unitary symmetry, $p(x_1)\to 0$  when $x_1\to 0$,
but $p(x_1=0)$ is non-zero in the case of chiral symmetry.
Since we are not able to calculate the sign of the 
parameters $x_i$ for a given sample, we plot in Fig.~\ref{figr2} 
the distribution of the \textit{absolute value} $|x_1|$. 
Dirichlet BC are imposed in the transversal direction.
For $E=0$, the distribution does not depend on the system size.
If $x_1$ possesses both positive and negative values, the distribution $p(x_1)$ 
is Gaussian with a mean value $\langle x_1\rangle=0$. This agrees with our data 
for the quasi-1d systems where we find $z_1=0$. 
Therefore, we conclude that the system possesses chiral symmetry.

However, the form of the distribution $p(x_1)$ 
changes qualitatively when the energy differs from zero. 
As is shown in Fig.~\ref{figr2}, already for $E=10^{-4}$ the distribution
$p(|x_1|)$ decreases to zero  when $|x_1|\to 0$. This confirms that the 
Jacobian given by Eq.~(\ref{jacoby}) contains also the repulsion term $\propto
\sin(2x)$. Consequently, the system changes the symmetry from chiral unitary
to unitary and all parameters $x_i$ become positive.
As shown in Fig.~\ref{figr2}, the distribution $p(x_1)$ converges to the
Wigner surmise $W_1$ when either $E$ or $L$ increases. Also, the distribution
of differences $x_2-x_1$ converges to the Wigner surmise $W_2$. This behavior
of $p(x_1)$ and $p(x_2-x_1)$ is typical for the unitary universality class.\cite{Pic91} 

\begin{figure}
\includegraphics[clip,width=0.4\textwidth]{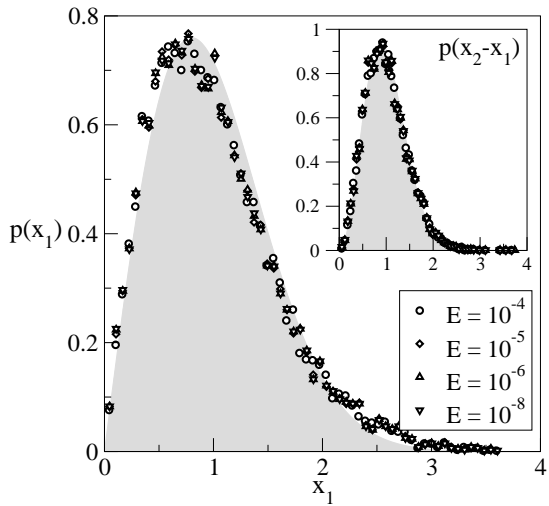}
\caption{The probability distribution $p(x_1)$ for a system with PBC at
  $E=0$. The system size is $65\times 66$ and $t_x=0.99$. $p(x_1)$ agrees well
  with the Wigner surmise for orthogonal and $p(x_1-x_2)$ (shown in the inset)
  for unitary ensembles.\cite{Pic91} 
}
\label{figr3}
\end{figure}

In case of $L$ odd and PBC in the transverse direction, the symmetry changes
to unitary and the critical point at $E=0$ disappears. 
The transfer matrix algorithm does not enable us to calculate the parameters
$x_i$ for $E=0$ and PBC due to the $k_z=0$ eigenmode of the transfer matrix
in unperturbed leads. This mode disappears either when $E\ne 0$ or when an
anisotropy in the hopping terms is applied. Using a small anisotropy 
in the $x$-direction, $t_x=0.99$, we confirmed that the statistics
of $p(x_1)$ and $p(x_2-x_1)$ follow the Wigner surmises also at the band center.
Fig.~\ref{figr3} shows the respective distributions $p(x_1)$ to be $W_1$ and
$p(x_2-x_1)$ is $W_2$. This is in contrast to the situation
with DBC where the distribution changes qualitatively on approaching $E=0$.

\begin{figure}
\includegraphics[clip,width=0.35\textwidth]{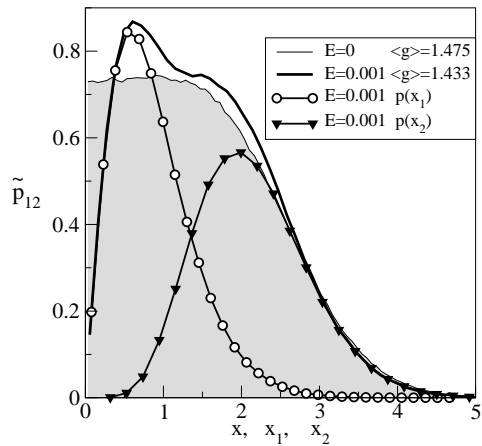}
\caption{
The probability distribution $\tilde{p}_{12}$, 
defined by Eq.~(\ref{pa}), for $E=0$ (shaded area) and for $E=0.001$. 
The size of the system is $66\times 66$. Dirichlet BC are used
in the transversal direction. Shown are also
distributions $p(x_1)$ and $p(x_2)$ for $E=0.001$. Note that $p\to 0$ when
$x\to 0$. This confirms that the system possesses different physical symmetry
for $E=0$ and $E\ne 0$, in agreement with \cite{MBF99}.
}
\label{figr4}
\end{figure}

\begin{figure}[b]
\includegraphics[clip,width=0.4\textwidth]{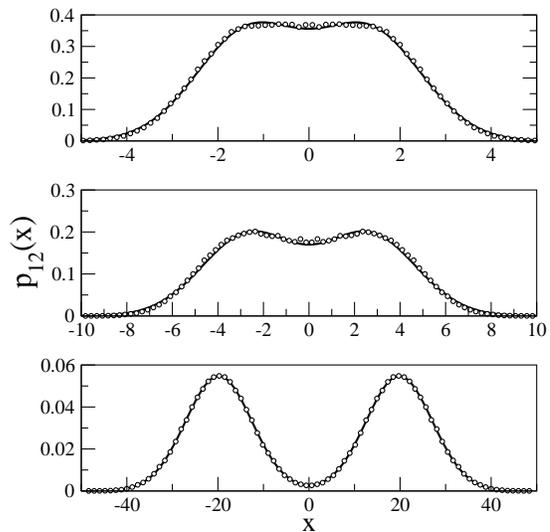}
\caption{
The distribution $p_{12}(x)$ for systems with even width $L=66$.
The length of the system is $L_z=66$ (top), $L_z=132$ (middle) and $L_z=1000$
(bottom). The solid line is the fit of $p_{12}(x)$, given by
Eq.~(\ref{2gauss}). 
}
\label{figr5}
\end{figure}

In case of $L$ even, the analysis is more difficult since we expect the mean
values of the first two parameters $x_1$ and $x_2$ to have the same absolute
value but with an opposite sign. So, we cannot distinguish between $|x_1|$ 
and $|x_2|$ in our analysis of a given sample. To overcome this problem, 
we calculate for $N_{\text{stat}}$ realizations the common  probability distribution,
\be\label{pa}
\tilde{p}_{12}(x)=\ds{\frac{1}{N_{\text{stat}}}}\sum_j^{N_{\text{stat}}}
\delta(x-|x_1|)+\delta(x-|x_2|),  
\ee
of the parameters $|x_1|$ and $|x_2|$ for square system $L\times L$ with $L=66$
and DBC in the transversal direction (Fig.~\ref{figr4}). 
We see that the probability $\tilde{p}_{12}(x)$ (shown by the shaded area) is
non-zero when $x\to 0$. We expect therefore that the distribution
\be
p_{12}(x)=
\ds{\frac{1}{N_{\text{stat}}}}\sum_j^{N_{\text{stat}}} \delta(x-x_1)+\delta(x-x_2) 
\ee
is of the form
\be\label{2gauss}
p_{12}(x)=\ds{\frac{1}{\sqrt{2\pi\sigma}}}
\left[\textrm{e}^{-\frac{(x-\langle x_1\rangle)^2}{2\sigma}}
+\textrm{e}^{-\frac{(x+\langle x_1\rangle)^2}{2\sigma}}\right].
\ee
This expectation is confirmed also by Fig.~\ref{figr5}, which shows how the
probability distribution changes when the system length increases.  

\begin{figure}[t]
\includegraphics[clip,width=0.39\textwidth]{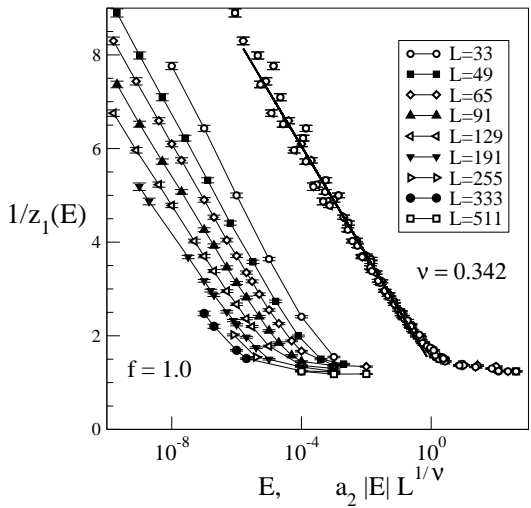}
\caption{The energy dependence of the spectrum of Lyapunov exponent
$z_1$ of the transfer matrix. Hard wall boundary conditions are imposed in the   
transversal direction and $f=1$.
The width of the system is given in the legend.  
The data scale to the universal curve on the right hand side, described by
Eq.~(\ref{scaling}), with critical exponent $\nu = 0.342$ and $a_2=0.05$.
}
\label{figr6}
\end{figure}

For square systems, we obtain the distribution shown in Fig.~\ref{figr4}
which, in the limit of $L_z/L\to\infty$ transforms into two Gaussian peaks.
For longer systems, $p(x_1,x_2)$ develops into two isolated Gaussian peaks
centered around the mean values, $\langle x_1\rangle=-\langle x_2\rangle$.
In analogy to the odd $L$ case, a non-zero energy breaks the chiral symmetry
also in the even $L$ situation.
A similar statistics  was observed also for PBC with small
anisotropy (not shown), which confirms the existence of the chiral symmetry
also for $L$ even and PBC. 

\begin{figure}[t]
\includegraphics[clip,width=0.4\textwidth]{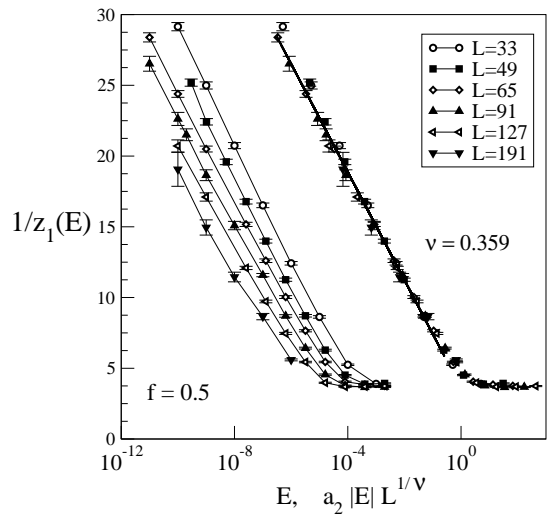}
\caption{The energy dependence of the spectrum of Lyapunov exponents
$z_1$ of the transfer matrix. Hard wall boundary conditions are imposed 
in the transversal direction and the flux strength is $f=0.5$.
The width of the system is given in the legend.  
The data scale to the universal curve (rhs) given by Eq.~(\ref{scaling}) with
$\nu = 0.359$ and $a_2=0.30$.
}
\label{figr7}
\end{figure}

We conclude that the random flux model with Dirichlet BC possesses at the band
center $E=0$ a chiral unitary symmetry. The spectrum of the Lyapunov exponents
is given by the relations 
\be\label{Levena}
z_i=c\times (-1)^{i+1}\,\textrm{Int}~\left[(i+1)/2\right]~~~~~(L~\textrm{even})
\ee
and
\be\label{Lodda}
z_i=
c\times (-1)^{i}\,\textrm{Int}~\left[i/2\right]~~~~~~~(L~\textrm{odd, DBC}),
\ee
in agreement with previous theoretical considerations.\cite{MBF99}

\section{Critical regime and exponent}\label{ce}
Since $z_1\equiv 0$ for $E=0$, $L$ odd and DBC in the transversal direction, 
the system is in the critical regime with a diverging correlation length 
\be
\xi\propto |E|^{-\nu}~~~~(\textrm{2d}).
\ee
To estimate the critical exponent $\nu$, we calculate $z_1$ as a function
of energy $E$ and of the system width $L$. We expect, in agreement with the
single parameter scaling,\cite{AALR79,MK81} that $z_1$ is a function
of the ratio $L/\xi(E)$ only.
As is shown in Fig.~\ref{figr6},
all numerical data can be fitted by the universal function 
\be\label{scaling}
z_1(E,L)=\displaystyle{\frac{a_1}{|\ln(a_2|E|L^{1/\nu})|}},
\ee
with three fitting parameters $a_1$, $a_2$, and $\nu$. 
From the scaling analysis we observed that
\be\label{exponent}
\nu= 0.35\pm 0.03.
\ee
More detailed information of the analysis is presented in Table~\ref{table2}.
To estimate the accuracy of our result, we repeated the scaling analysis with
reduced input data sets. 
A similar value of the critical exponent was obtained also for systems 
with weaker magnetic flux disorder $f=0.5$ (see Fig.~\ref{figr7}).
Due to the smaller values of $z_1$, we have to simulate much longer quasi-1d
systems in order to get data with reasonable accuracy. 

\begin{table}
\begin{tabular}{lllr}
$L_{\text{min}}$  &   $z_{1{\text{max}}}$   &    $\nu$   & $F_{\text{min}}/N_{\text{data}}$\\
\hline
\multicolumn{3}{c}{$f=1.0$}\\
\hline
33    &    0.62    &  0.342   &  20/69\\
33    &    0.52    &  0.336   &  10/58\\
65    &    0.62    &  0.348   &  13/55 \\
65    &    0.52    &  0.342   &   7/49\\
91    &    0.62    &  0.368   &   7/39\\
91    &    0.52    &  0.355   &   4/35\\
91    &    0.45    &   0.341  &   3/30\\
91    &   0.40     &  0.329   &   2/25\\
\hline
\multicolumn{3}{c}{$f=0.5$}\\
\hline
33    &     0.15   &    0.359 &    4/36\\
33    &     0.20   &    0.384  &   11/42\\
33    &    0.25    &    0.372  &   26/45\\
\hline
\end{tabular}
\caption{Numerical estimate of the critical exponent $\nu$ for two different
  strengths of the random flux amplitudes, $f=1.0$ and $f=0.5$. Only data for 
  $L>L_{\text{min}}$ and with $z_1<z_{1{\text{max}}}$ are considered in the
  scaling analysis.  
  $F_{\text{min}}$ is obtained from the minimum of the fitting function,
  $N_{\text{data}}$ is the number of data. The accuracy of the critical
  exponent in each fitting procedure is of the order of $10^{-3}$.}
\label{table2}
\end{table}

Although the calculated values of $\nu$ for $f=0.5$ differ slightly from those
obtained for $f=1.0$, we do not interprete this difference as a non-universality
of the critical exponent.\cite{ERS01}  Rather we assume that this difference
is due to the limited accuracy of our numerical data and/or fitting procedure. 
Indeed, as shown in Table~\ref{table2}, the estimated value of the critical
exponent depends on the choice of the input ensemble defined by $z_{\text{1max}}$
and $L_{\text{min}}$, and decreases slightly when larger values of $z_1$ are
excluded. 

We did not find any crossover from a power-law to a more complicated $E$
dependence of the localization length for 2d as proposed in
Ref.~\onlinecite{FC00} and  discussed in Refs.~\onlinecite{ERS01} and
\onlinecite{ER04}. Our numerical data cannot be fitted to 
the one parameter scaling function  $z_1(E,L)=z_1(L/\xi)$ with a localization
length $\xi\propto \exp\sqrt{\ln (E_0/E)}$.\cite{FC00} 
Since we analyze a very narrow energy interval around the band center 
(as small as $|E|\sim 10^{-10}$), we do not expect that the crossover from the
observed power-law to the proposed logarithmic energy dependence of the
localization length exists in our situation. 

Since $z_1$ determines the localization length of the quasi-1d system,
$\xi=2/z_1$, we see from Eq.~(\ref{scaling}) that the localization length
diverges as
\be
\xi_L(E)\propto L\times |\ln (a_2|E|L^{1/\nu})|~~~~~~{\text{(Q1D),}}~E\to 0
\ee
for a given system width $L$. A logarithmic divergence is typical for
Anderson bond disordered models.\cite{TC76,Mar88}

\begin{figure}[t]
\includegraphics[clip,width=0.4\textwidth]{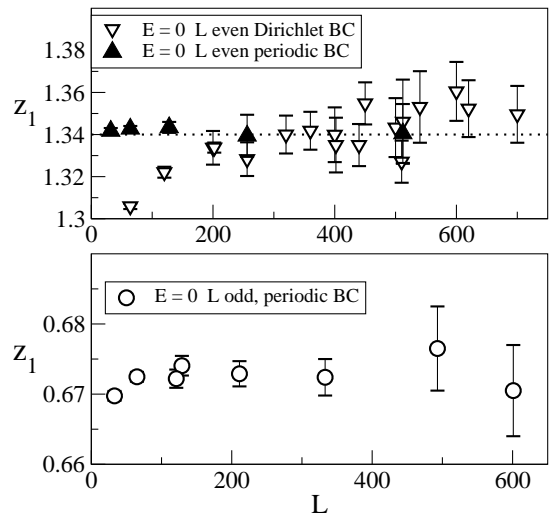}
\caption{
The smallest Lyapunov exponent $z_1$ as a function of the system width $L$
for $L$ even (top) and $L$ odd (bottom). The data confirm
that $z_1$ does not depend on the system width. This either implicates the
existence of a critical point at the band center in all three cases or a
finite size effect due to the limited system size.
}
\label{figr8}
\end{figure}

While the existence of the critical state at $E=0$ for $L$ even is commonly 
accepted,\cite{Fur99} we do not expect the same for $L$ odd and PBC, since the 
system possesses unitary symmetry in this case.
To describe the property of the $E=0$ state, we plot in 
Fig.~\ref{figr8} the $L$ dependence of the smallest LE $z_1$ for
$L$ even (both Dirichlet and periodic BC) and $L$ odd (PBC). 
In all three situations, we do not observe any $L$ dependence of the
smallest LE $z_1$. We believe that this indicates that the localization
lengths considerably exceeds the available system sizes so that no final
conclusions can be reached.  

\begin{figure}[b]
\includegraphics[clip,width=0.38\textwidth]{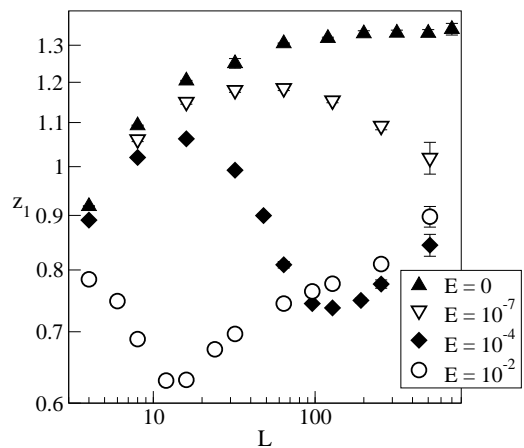}
\caption{The $L$ dependence of the smallest Lyapunov exponent $z_1$ for $L$
  even and for various values of the energy $E$. The non-monotonous $L$
  dependence disables the scaling analysis in this case. Our data are
  consistent with Fig.~1 of Ref.~\onlinecite{Cer01}.
}
\label{figr9}
\end{figure}

The scaling analysis is very difficult in the case of $L$ even. 
As is shown in Fig.~\ref{figr9}, the $L$-dependence of $z_1$ is highly
non-trivial for non-zero energies in accordance with Ref.~\onlinecite{Cer01}. 
The scaling seems to work only in the limit of $L\to\infty$. 
We disagree on the observation\cite{Cer01} that the LEs at $E=0$
do not come in degenerate pairs. In contrast we find the difference between
the two LEs to be smaller than the accuracy of our calculations.

The criticality of the $E=0$ state for the Dirichlet BC will be supported also
by the size dependence of the mean conductance, discussed in the next Section.

\begin{figure}
\includegraphics[clip,width=0.48\textwidth]{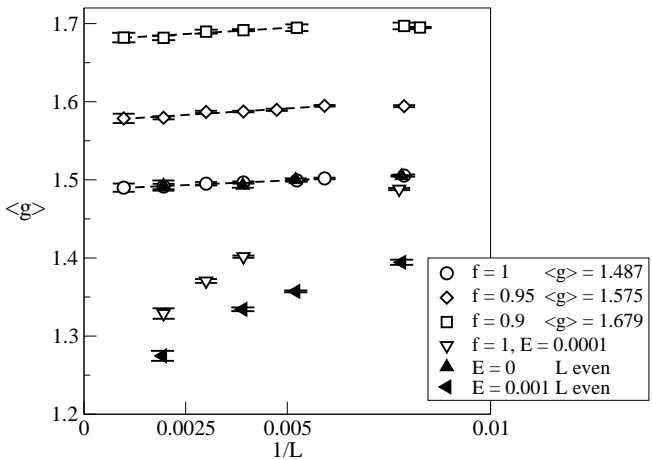}
\caption{
The critical value of the mean conductance $\langle g_c\rangle$ as a function
of the system's size $L\times L$ with $L$ odd (open symbols) and $L$ even
(full symbols) at $E=0$ with DBC, and various strengths of the random field 
$f$. The data show that the critical conductance does not depend on the
parity of $L$, but depends on $f$.
For completeness, we add also data for $E=0.0001$ ($L$ odd) 
and $E=0.001$ ($L$ even) to show that the conductance
decreases with $L$ when $E\ne 0$, indicating that the system is in the
localized regime in the limit of $L\to\infty$. 
}
\label{figr10}
\end{figure}

\begin{figure}[b]
\includegraphics[clip,width=0.35\textwidth]{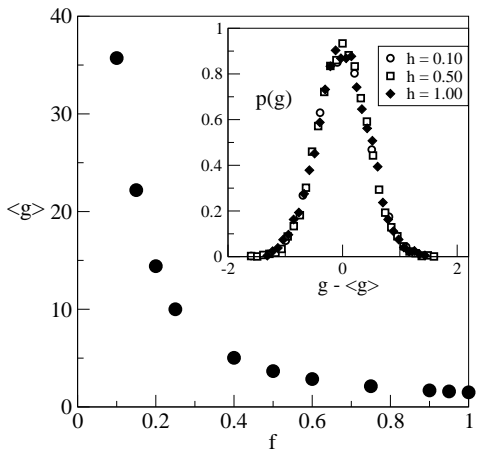}
\caption{
The critical value of the mean conductance $\langle g_c\rangle$
for square samples $257\times 257$ at $E=0$ and various values 
of the random field $f$. Dirichlet BC are used in the transversal direction.
The inset shows the probability distribution
$p(g-\langle g\rangle)$ for three different values of $f$. The width of the
distribution var\,$g \approx 0.187$ does not depend on $f$.}
\label{figr11}
\end{figure}

\section{Conductance}\label{cond}
Figure~\ref{figr10} shows the size dependence of the sample averaged critical
conductance $\langle g_c\rangle$ for square systems $L\times L$ ($L$ odd) and
three values of the randomness strength $f$. The energy is $E=0$ and Dirichlet
BC are considered.  
Our data confirm that $\langle g\rangle$ converges to an $L$ independent   
critical value $\langle g_c\rangle$ which, however, does depend on the
strength of the randomness $f$. For the largest possible disorder $f=1$
we obtain $\langle g_c\rangle=1.49$, a value larger than the 2d symplectic 
case.\cite{MS06} Figure~\ref{figr11} shows the 
$f$ dependence of the mean conductance for squares of size $257\times 257$.
It also shows that the variance var\,$g = \langle g^2\rangle - \langle g\rangle^2$
is universal and independent on $f$. 
We find a value var\,$g \approx 0.187$ which is in agreement with those
obtained earlier by Ohtsuki \textit{et al.}\cite{OSO93} and
Furusaki.\cite{Fur99} 
We observe, however, an increase of var~$g$ for very small $f$, which can be
explained by finite size effects due to a large mean free path.

We also plot in Fig.~\ref{figr10} the size dependence of the mean
conductance for squares with even $L$. Within the obtained accuracy, $\langle
g_c\rangle$ does neither depend on the parity of $L$ nor on the boundary
conditions
in agreement with Ref.~\onlinecite{Fur99}.
Contrary to the band center, the conductance decreases always with increasing
system size whenever the energy lies outside the band center.

We also analyzed the length dependence of the mean conductance 
$\langle g\rangle$ and of the mean of the logarithm of the conductance
$\langle\ln g\rangle$ 
for systems with hard wall transversal boundary conditions and $E=0$.
Since $\langle x_1\rangle\equiv 0$ in this case, we expect that the 
system possesses an infinite localization length also in the quasi-1d
limit.\cite{MBF99}  
Therefore, the mean conductance $\langle g\rangle$ should not decrease
exponentially when the system length increases. 

\begin{figure}
\includegraphics[clip,width=0.4\textwidth]{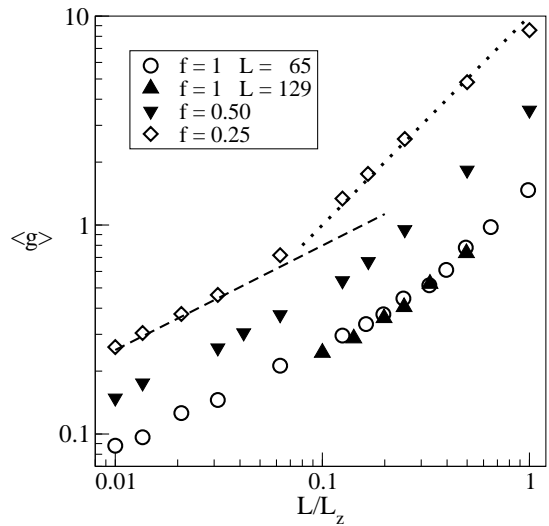}
\caption{
The length dependence of the mean conductance $\langle g\rangle$
for Q1D systems. The systems width is $L=65$, the energy $E=0$.
One clearly sees the crossover from the $1/L$ (dotted line) 
to $1/\sqrt{L_z}$(dashed line) dependence
predicted by Ref.~\onlinecite{MBF99}.
}
\label{figr12}
\end{figure}

Our results shown in Fig.~\ref{figr12} confirm the relations predicted
theoretically\cite{MBF99} 
\be
\langle g\rangle=L\ell/L_z
\ee
and 
\be\langle g\rangle = \sqrt{2L\ell/(\pi L_z)}\ee
in the limit of $\langle g\rangle\approx 1$  
and $\langle g\rangle \ll 1$, respectively.

\section{Summary}\label{concl}
We investigated two-dimensional electron systems with static random magnetic
flux and showed numerically that the transport properties  
depend on the parity of the system's width $L$ and on the transverse
boundary conditions. For Dirichlet boundary conditions, we confirmed by the
analysis of the statistical properties of the quantities $x$, which
parameterize the eigenvalues of the transmission matrix, that the system
possesses chiral unitary symmetry at the band center. 
The chirality exists in case of Dirichlet boundary conditions for both $L$ odd
and even, and for periodic BC for $L$ even only.
But the chirality is always broken when the energy of the electron is non-zero.

In case of chiral unitary symmetry, the 2d
system with random magnetic flux possesses a critical point at the band
center. We found that the localization length diverges $\propto |E|^{-\nu}$
when $E\to 0$ and calculated the critical exponent $\nu\approx 0.35$ for $L$
odd and Dirichlet BC. Our data do not confirm the existence of the crossover
from the power-law to the logarithmic energy dependence of $\xi$ predicted by
Ref~\onlinecite{FC00}.  

We also calculated the critical conductance of 2d systems. At the band center, 
the mean conductance $\langle g\rangle$ converges to a size-independent
critical value for both $L$ odd and $L$ even. Although the critical
conductance does depend on the strength of the randomness, the fluctuations of
the conductance appear to be universal. For non-zero energy, the mean 
conductance decreases with the system size, indicating a localized regime.
Finally, for the quasi-1d systems with odd system width and Dirichlet BC, 
we confirmed the non-trivial length dependence of the mean conductance, 
proposed theoretically in Ref.~\onlinecite{MBF99}.

\section{Acknowledgment}
PM thanks grant APVV project No.~51-003505, VEGA project No.~2/6069/26,
and PTB for hospitality.


\begin{thebibliography}{10}

\bibitem{LF81}
P.~A. Lee and D.~S. Fisher, Phys. Rev. Lett. {\bf 47},  882  (1981).

\bibitem{SN93}
T. Sugiyama and N. Nagaosa, Phys. Rev. Lett. {\bf 70},  1980  (1993).

\bibitem{Fur99}
A. Furusaki, Phys. Rev. Lett. {\bf 82},  604  (1999).

\bibitem{Cer00}
V.~Z. Cerovski, Phys. Rev. B {\bf 62},  12775  (2000).

\bibitem{Cer01}
V.~Z. Cerovski, Phys. Rev. B {\bf 64},  161101(R)  (2001).

\bibitem{ERS01}
A. Eilmes, R.~A. R\"omer, and M. Schreiber, Physica B {\bf 296},  46  (2001).

\bibitem{EK03}
S.~N. Evangelou and D.~E. Katsanos, J. Phys. A: Math. Gen. {\bf 36},  3237
  (2003).

\bibitem{GT04}
A.~M. Garc\'{i}a-Garc\'{i}a and K. Takahashi, Nucl. Phys. {\bf B700},  361
  (2004).

\bibitem{ER04}
A. Eilmes and R.~A. R\"omer, phys. stat. sol. (b) {\bf 241},  2079  (2004).

\bibitem{GC06}
A.~M. Garc\'{i}a-Garc\'{i}a and E. Cuevas, Phys. Rev. B {\bf 74},  113101
  (2006).

\bibitem{KZ92}
V. Kalmeyer and S.-C. Zhang, Phys. Rev. B {\bf 46},  9889  (1992).

\bibitem{HLR93}
B.~I. Halperin, P.~A. Lee, and N. Read, Phys.\ Rev.\ B {\bf 47},  7312  (1993).

\bibitem{LC94}
D.~K.~K. Lee and J.~T. Chalker, Phys. Rev. Lett. {\bf 72},  1510  (1994).

\bibitem{NL90}
N. Nagaosa and P.~A. Lee, Phys. Rev. Lett. {\bf 64},  2450  (1990).

\bibitem{AI92}
B.~L. Altshuler and L.~B. Ioffe, Phys. Rev. Lett. {\bf 69},  2979  (1992).

\bibitem{SV93}
E.~V. Shuryak and J.~J.~M. Verbaarschot, Nucl. Phys. {\bf A560},  306  (1993).

\bibitem{AV00}
A.~M. Garc\'{i}a-Garc\'{i}a and J.~J.~M. Verbaarschot, Nucl. Phys. {\bf B586},
  668  (2000).

\bibitem{GO04}
A.~M. Garc\'{i}a-Garc\'{i}a and J.~C. Osborn, Phys. Rev. Lett. {\bf 93},
  132002  (2004).

\bibitem{GO07}
A.~M. Garc\'{i}a-Garc\'{i}a and J.~C. Osborn, Phys. Rev. D {\bf 75},  034503
  (2007).

\bibitem{MBF99}
C. Mudry, P.~W. Brouwer, and A. Furusaki, Phys. Rev. B {\bf 59},  13221
  (1999).

\bibitem{BSK98}
M. Batsch, L. Schweitzer, and B. Kramer, Physica B {\bf 249-251},  792  (1998).

\bibitem{PS99}
H. Potempa and L. Schweitzer, Ann. Phys. (Leipzig) {\bf 8},  SI\symbol{45}209
  (1999).

\bibitem{AMW94}
A.~G. Aronov, A.~D. Mirlin, and P. W\"olfle, Phys. Rev. B {\bf 49},  16609
  (1994).

\bibitem{ITA94}
M. Inui, S.~A. Trugman, and E. Abrahams, Phys. Rev. B {\bf 49},  3190  (1994).

\bibitem{GW91}
R. Gade and F. Wegner, Nucl. Phys. {\bf B360},  213  (1991).

\bibitem{Gad93}
R. Gade, Nucl. Phys. {\bf B398},  499  (1993).

\bibitem{MW96}
J. Miller and J. Wang, Phys. Rev. Lett. {\bf 76},  1461  (1996).

\bibitem{BMSA98}
P.~W. Brouwer, C. Mudry, B.~D. Simons, and A. Altland, Phys. Rev. Lett. {\bf
  81},  862  (1998).

\bibitem{AS99}
A. Altland and B.~D. Simons, J. Phys. A: Math. Gen. {\bf 32},  L353  (1999).

\bibitem{FC00}
M. Fabrizio and C. Castelliani, Nucl.\ Phys.\ B {\bf 583},  542  (2000).

\bibitem{ES81}
E.~N. Economou and C.~M. Soukoulis, Phys. Rev. Lett. {\bf 46},  618  (1981).

\bibitem{PZIS90}
J.-L. Pichard, N. Zanon, Y. Imry, and A.~D. Stone, J. Phys. France {\bf 51},
  587  (1990).

\bibitem{Mar95a}
P. Marko\v{s}, J. Phys.: Condens. Matter {\bf 7},  8361  (1995).

\bibitem{Ose68}
V.~I. Oseledec, Trans. Moscow Math. Soc. {\bf 19},  197  (1968).

\bibitem{PS81}
J.-L. Pichard and G. Sarma, J. Phys. C: Solid State Phys. {\bf 14},  L127
  (1981).
J.-L. Pichard and G. Sarma, J. Phys. C: Solid State Phys. {\bf 14},  L617
  (1981).

\bibitem{MK81}
A. MacKinnon and B. Kramer, Phys. Rev. Lett. {\bf 47},  1546  (1981).

\bibitem{Pic91}
J.-L. Pichard,  in {\em Quantum Coherence in Mesoscopic Systems}, Vol.~254 of
  {\em Nato ASI}, edited by B. Kramer (Plenum Press, New York, 1991), pp.\
  369--399.

\bibitem{Dor82a}
O.~N. Dorokhov, JETP Lett. {\bf 36},  318  (1982).

\bibitem{MPK88}
P.~A. Mello, P. Pereyra, and N. Kumar, Ann. Phys. (N.Y.) {\bf 181},  290
  (1988).

\bibitem{AALR79}
E. Abrahams, P.~W. Anderson, D.~C. Licciardello, and T.~V. Ramakrishnan, Phys.
  Rev. Lett. {\bf 42},  673  (1979).

\bibitem{TC76}
G. Theodorou and M.~H. Cohen, Phys. Rev. B {\bf 13},  4597  (1976).

\bibitem{Mar88}
P. Marko\v{s}, Z. Physik B {\bf 73},  17  (1988).

\bibitem{MS06}
P. Marko\v{s} and L. Schweitzer, J. Phys. A {\bf 39},  3221  (2006).

\bibitem{OSO93}
T. Ohtsuki, K. Slevin, and Y. Ono, J. Phys. Soc. Japan. {\bf 62},  3979
  (1993).

\end{thebibliography}


\end{document}